\documentclass{ws-ijmpc}

\begin{document}

%\markboth{F. Author \& S. Author (authors' names)}
%{Instructions for typing manuscripts (paper's title)}

\catchline{}{}{}{}{}
%%%%%%%%%%%%%%%%%%%%%%%%%%%%%%%%%%%%%%%%%%%%%%%%%%%%%%%%%%%%%%%%%%%%

\title{Improved phase-field-based lattice Boltzmann models with a filtered collision operator}

\author{Hiroshi Otomo, Raoyang Zhang, Hudong Chen}

%\footnote{
%Typeset names in 8~pt Roman, upper and lower case. Use the footnote to indicate the
%resent or permanent address of the author.}}

\address{Dassault Syst\'{e}mes, 55 Network Drive, Burlington, MA 01803, USA \\ Hiroshi.Otomo@3ds.com}
%\\
%first\_author@domain\_name}

\maketitle

\begin{history}
\received{Day Month Year}
\revised{Day Month Year}
\end{history}

\begin{abstract}
In this study, a phase-field lattice Boltzmann model based on the Allen-Cahn equation with a filtered collision operator and high-order corrections in the equilibrium distribution functions is presented. 
Here we show that in addition to producing numerical results consistent with prior numerical methods, analytic solutions, and experiments  with the density ratio of 1000,  previous numerical deficiencies are resolved.
Specifically, the new model is characterized by robustness at low viscosity, accurate prediction of shear stress at interfaces, and  removal of artificial dense bubbles and rarefied droplets, etc.
\end{abstract}

\ccode{PACS Nos.: 11.25.Hf, 123.1K}

%%%%%%%%%%%%%%%%%%% section start %%%%%%

\section{Introduction}

The modified Allen-Cahn (AC) equation, in which curvature driven dynamics is eliminated \cite{2007_Sun,2011_Chiu}, is widely studied as an efficient interface-tracking solver in multiphase flow with high density ratio due to its robustness and handiness of the conserved second-order partial differential equation.

In previous studies\cite{2016_Fakhari,2017_Liang}, a second set of the distribution functions is introduced to solve the AC equation for the interface dynamics, which is coupled with the hydrodynamic lattice Boltzmann (LB) based momentum solver.
In order to improve stability at low viscosity and accuracy around interfaces,
the multiple relaxation time (MRT) scheme and the biased difference scheme for
the gradient of the order parameter were implemented \cite{2016_Fakhari}. 
The MRT scheme, however, suffers from high computational costs and a large number of model parameters.
Also, due to using the biased difference scheme, the requirement of information
from sites next to the nearest neighbors causes difficulties in near boundary
regions, which may involve nontrivial extrapolation, complex code vectorization
and parallelization, and additional computational cost.

In the present study, our LB model is formulated with a filtered collision
operator\cite{2006_Chen,2006_Zhang,2006_Latt,2006_Shan,2013_Chen,2014_Chen} with
a single relaxation time. The central difference scheme, which only requires
information at nearest neighbor sites, is applied because of its high computational
efficiency. 
It is known that the truncation error of the central difference scheme may cause inaccurate shear stress on interfaces \cite{2017_Fakhari}. 
Also, the original AC equation may inevitably produce artificial dense bubbles and rarefied droplets.
We propose a single solution to both of these unphysical effects and present numerical results from our improved model where these undesirable features are mitigated.
All validation cases are conducted with a single source code containing our newly proposed algorithm.

\section{Lattice Boltzmann models}
\label{LB_model_pot_issue}

Two lattice Boltzmann (LB) equations, one for the order parameter $\phi$ and the other for hydrodynamic quantities such as pressure $P$ and momentum $\rho \vec{u}$ are solved. 
For both equations, the D3Q19 or D2Q9 lattice model is adopted with index $i \in \left\{ 1, \cdots 19 \right\}$ or $i \in \left\{ 1, \cdots   9 \right\}$.

Formulation of the LB equation for $\phi$ follows the previous study\cite{2016_Fakhari}, 
\begin{equation}
\label{eq:LB_phi_filter_col}
 {h}_{i} \left( \vec{x}+\vec{c}_{i} \Delta t, t+\Delta t \right) =
 {h}_{i} \left( \vec{x}, t \right) - \frac{{h}_{i} - {h}^{eq}_{i}}{\left(M/T\right)+ 0.5} \vert_{\vec{x}, t},
\end{equation}
where $M$ is the mobility and ${h}^{eq}_{i}$ is the equilibrium state defined as,
\begin{eqnarray}
{h}^{eq}_{i} &=& \phi \Gamma_i + \theta w_i \left( \vec{c}_i \cdot \vec{n} \right), \\
\Gamma_i  &=& w_i \left\{ 1+ \frac{\vec{c}_i \cdot \vec{u} }{T} + \frac{\left(  \vec{c}_i \cdot \vec{u} \right)^2}{2 T^2} - \frac{ \vec{u}^2}{2T} \right\}, \\
\theta &=& \frac{M}{T} \left\{ \frac{1- 4 \left( \phi - 0.5 \right)^2 }{W} \right\}. \label{theta_phai_LBeq}
\end{eqnarray}
Here $W$ is  the interface thickness and $ \vec{n}$ is the unit vector normal to the interface calculated by $\vec{\nabla} \phi / \left(  |\vec{\nabla} \phi|  + \epsilon \right)$ where $\epsilon$ is a tiny parameter such as $1.e-10$ in order to avoid division by zero. The value of $\phi$ is evaluated by $\sum_i {h}_i$.

Formulation of the hydrodynamic LB equation is based on the previous study\cite{2016_Fakhari}. 
One with the BGK collision operator is,
\begin{align}
\label{BGK_LBeq_hydroeq}
\bar{g}_{i} \left( \vec{x}+\vec{c}_{i} \Delta t, t+\Delta t \right) =
 \bar{g}_{i} \left( \vec{x}, t \right) - \frac{\bar{g}_{i} - \bar{g}^{eq}_{i}}{\tau_{mix}} \vert_{\vec{x}, t} + K_i \left( \vec{x}, t \right),
\end{align}
where
\begin{eqnarray}
\bar{g}^{eq}_{i}&=& \rho \Gamma_i + w_i \left( \frac{P}{T} - \rho \right)- \frac{K_i}{2},  \\
K_i &=& \left\{ \left(  \Gamma_i - w_i \right)  \rho_{dif} + \frac{\Gamma_i \mu_{chm} }{T}  \right\} \left( \vec{c}_{i} - \vec{u}  \right) \cdot \vec{\nabla} \phi + \Gamma_i \frac{\left( \vec{c}_{i} - \vec{u}  \right) \cdot  \vec{F}_{ex} }{T}.
\end{eqnarray}
Here $K_i$ is a force term responsible for phase separation and the external force $\vec{F}_{ex}$.
$\rho_{dif}$ is difference between the light and heavy characteristic density $\rho$.
$\mu_{chm}$ is the chemical potential defined as,
\begin{eqnarray}
\mu_{chm}=\frac{48 \sigma}{W} \phi \left(  \phi -1 \right) \left( \phi -0.5 \right) - \frac{3 \sigma W}{2} \vec{\nabla}^2 \phi,
\end{eqnarray}
where $\sigma$ is the surface tension.

The relaxation time $\tau_{mix}$ is approximated using the harmonic interpolation,
\begin{equation}
\label{tau_interpolate}
1/ \tau_{mix} = \left( 1 /\tau_{air} \right) + \phi \left\{ \left( 1/\tau_{water} \right) - \left(1/\tau_{air} \right) \right\},
\end{equation}
with relaxation times of water and air, $\tau_{water}$ and $\tau_{air}$, which
correspond to their kinematic viscosities, $\nu_{water}$ and $\nu_{air}$.

The right hand side  in Eq.~(\ref{BGK_LBeq_hydroeq}) is  filtered as,
\begin{equation}
\label{eq:LB_filter_col}
 \bar{g}_{i} \left( \vec{x}+\vec{c}_{i} \Delta t, t+\Delta t \right) =
 \bar{G}_{i}^{eq} + \left( 1 - \frac{1}{ \tau_{mix} }  \right) \Phi_i : \Pi,
\end{equation}
where $\Phi_i$ is a filtered operator that uses Hermite polynomials and $\Pi$ is the nonequilibrium moments of the momentum flux,
\begin{eqnarray}
\Phi_i = \frac{w_i}{2 T^2} \left( \vec{c}_{i} \vec{c}_{i} - T I \right), \label{momentum_flux0} \\
\Pi = \sum_{l}  \vec{c}_{l} \vec{c}_{l} \left(   \bar{g}_{l} - \bar{G}_{l}^{neq} \right). \label{momentum_flux}
\end{eqnarray}
Here $I$ is the identity matrix.
The equilibrium and nonequilibrium parts $\bar{G}_{i}^{eq}$ and $\bar{G}_{i}^{neq}$ are naturally determined via correspondence with Eq.~(\ref{BGK_LBeq_hydroeq}) and $\tau_{mix}$ dependence.
More details of filtered collision procedure can be found in previous studies \cite{2006_Chen,2006_Zhang,2006_Latt,2006_Shan,2013_Chen,2014_Chen,2016_Otomo,2018_Otomo}. 
Projection from the state-space to moment-space is performed in Eq.~(\ref{momentum_flux}) only for 9 moments in the case of D3Q19.
Also, since the spatially dependent relaxation time $\tau_{mix}$ is factored out from the projected term in Eq.~(\ref{eq:LB_filter_col}), the calculation of $\Phi_i : \Pi$ can be simplified down to the multiplication between a $19\times19$ matrix \footnote{Simply following Eq.~(\ref{momentum_flux0}) and Eq.~(\ref{momentum_flux}), we can decompose this $19\times19$ matrix to the multiplication between a $19\times3$ matrix and a  $3 \times19$ matrix so that the operation count is saved further. } and a $1 \times 19$ matrix.
On the other hand, in the MRT scheme the projection is performed for the 19 moments on D3Q19 via the multiplication of the matrix $M^{-1}DM$ where $M$ is the $19\times19$ conversion matrix from the state-space to the moment space and $D$ is the 19th-rank diagonal matrix involving the relaxation time. This multiplication of matrices is known to cause deterioration of computational efficiency.
As a result, the filtering discussed in the present study can possess much higher computational efficiency compared to the MRT scheme.

After Eq.~(\ref{eq:LB_filter_col}) is solved, pressure and momentum are evaluated by $T \sum_i \bar{g}_i+\left(T \rho_{dif}/2 \right) \vec{u} \cdot \vec{\nabla} \phi$ and $\sum_i \vec{c}_i \bar{g}_i + \left( \mu_{chm} \vec{\nabla} \phi +\vec{F}_{ex} \right)/2$, respectively.

In this study, the gradient and Laplacian of $\phi$, that are used for calculation of $\vec{n}$, $P$, $\rho \vec{u}$, and $K_i$, are approximated with the central difference (CD) scheme,
\begin{equation}
\vec{\nabla} \phi =\frac{ \sum_i  \left\{ \phi \left( \vec{x} + \vec{c}_i \right) -\phi \left( \vec{x} - \vec{c}_i \right) \right\} \vec{c}_i w_i}{2T},
\: \: \: \: \: \:
\vec{\nabla}^2 \phi = \frac{2 \sum_i \left\{ \phi \left( \vec{x} + \vec{c}_i \right) -\phi \left( \vec{x} \right) \right\} w_i}{T}. \nonumber
\end{equation}
This discretization scheme requires information only from the nearest neighbor sites, whereas existing implementations utilizing biased differencing use next-nearest neighbor sites.
For all results shown in this manuscript, unless specifically mentioned, the density ratio $\rho_{ratio}= \rho_{water}/ \rho_{air}$ is 1000, consistent with air - water mixtures. The dynamic viscosity ratio $\mu_{water}/\mu_{air}$ is denoted as $\mu_{ratio}$.

\section{Pathological cases}

In spite of high computational efficiency, the LB models described in the previous section are inadequate for some benchmark cases.
In this section, a few simple cases are chosen as pathological cases and the remedies are proposed.

\

\begin{figure}[htbp]
  \begin{center}
    \begin{tabular}{c}
      \begin{minipage}{0.33\hsize}
        \begin{center}
          \includegraphics[clip, width=4.2 cm, height = 3.5cm]{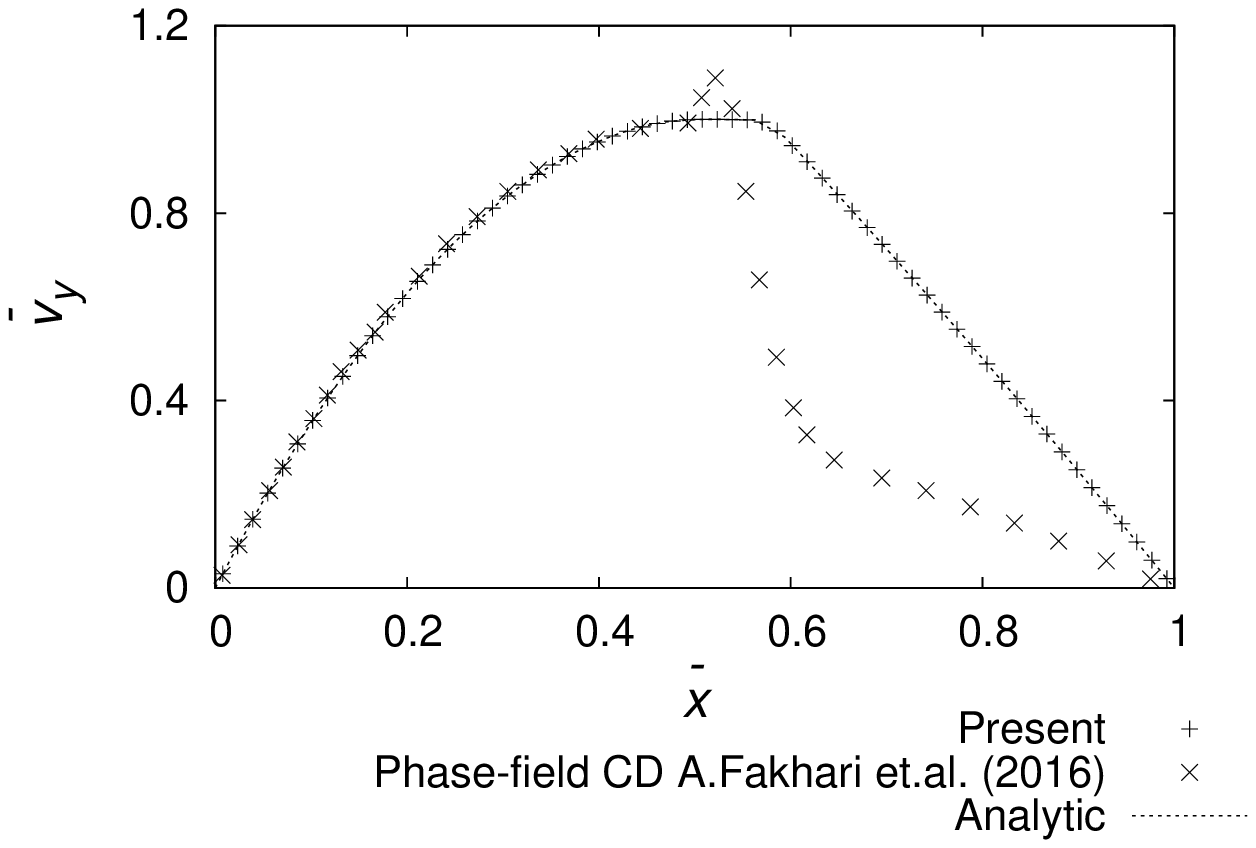}
        \end{center}
      \end{minipage}
      \begin{minipage}{0.33\hsize}
        \begin{center}
          \includegraphics[clip, width=4.2 cm, height = 3.5cm]{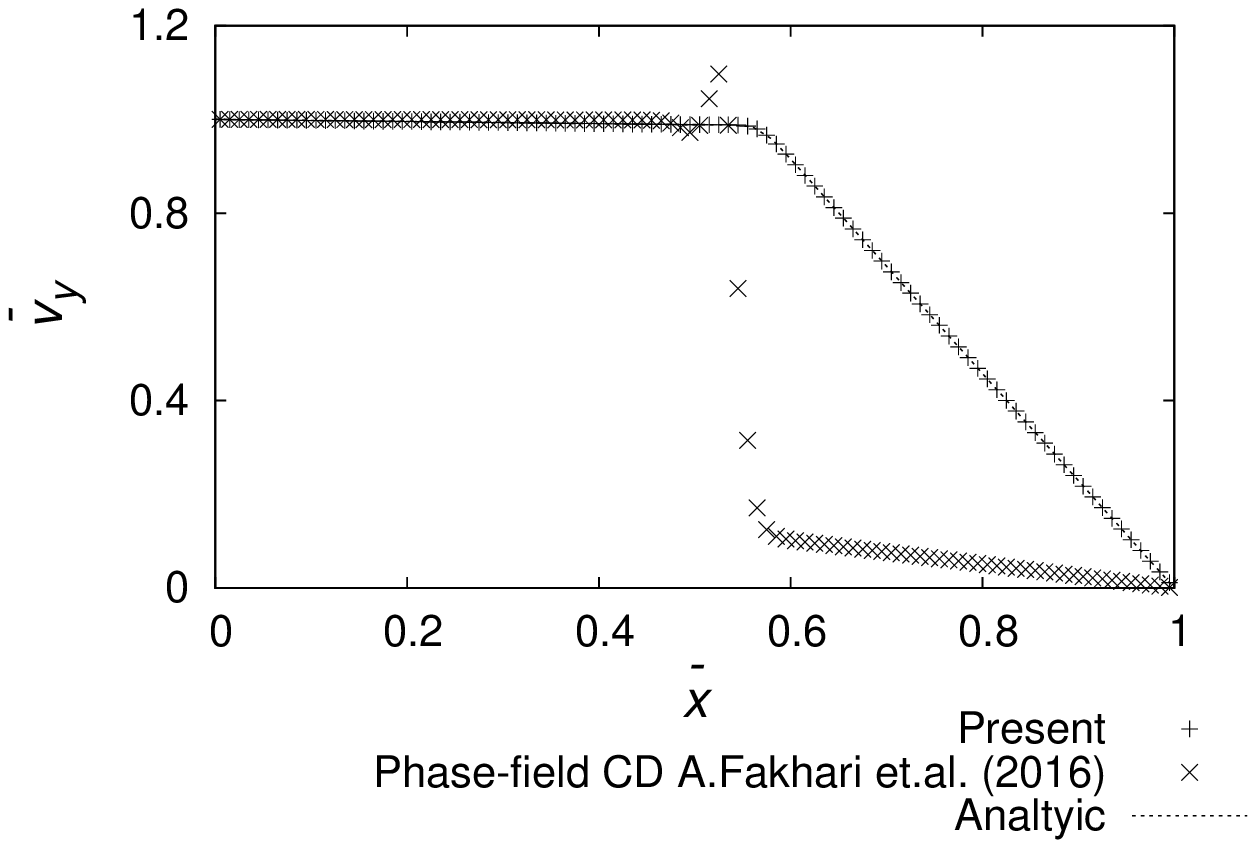}
        \end{center}
      \end{minipage}
       \begin{minipage}{0.33\hsize}
        \begin{center}
          \includegraphics[clip, width=4.2 cm, height = 3.5cm]{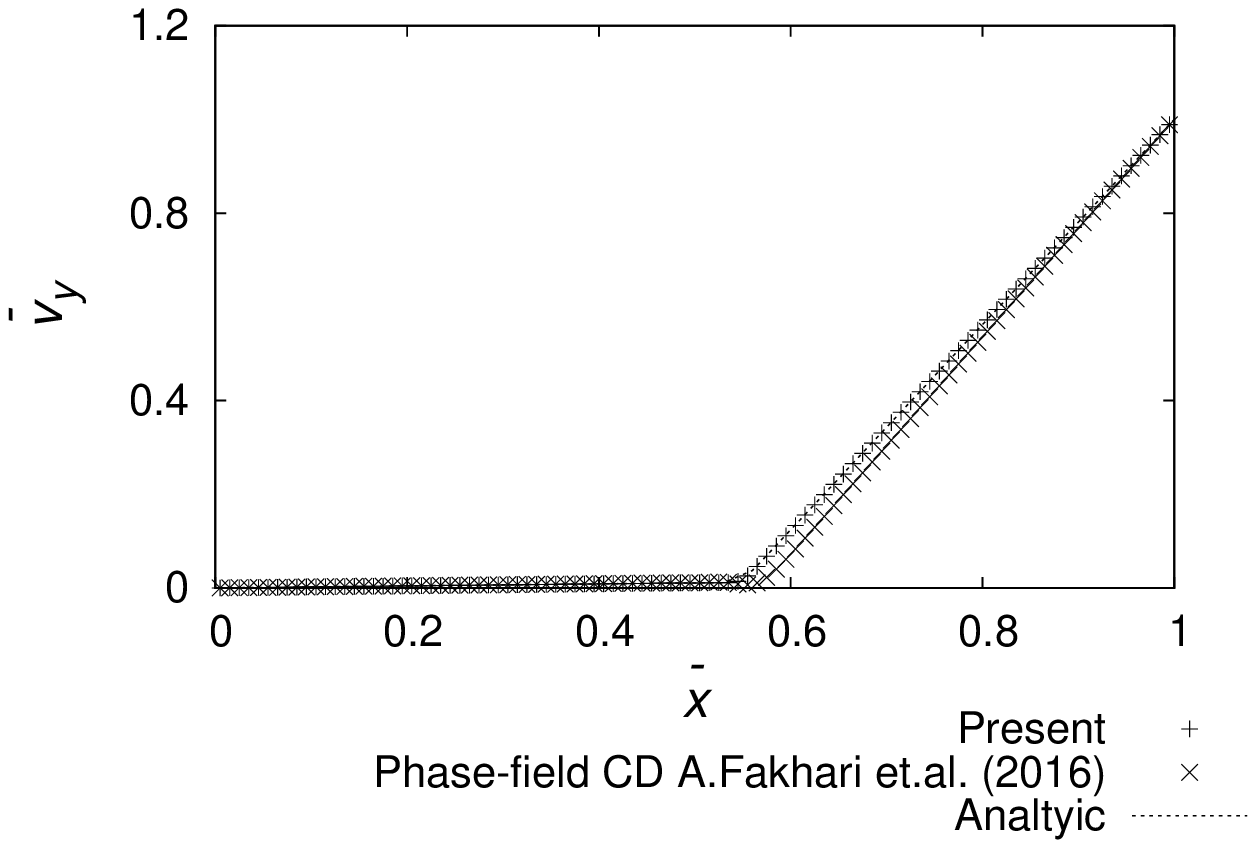}
        \end{center}
      \end{minipage}
    \end{tabular}
    \caption{ Velocity profile of multi-phase Poiseuille (left) and Couette (center and right) flow normalized by maximum velocity in the analytic solution. 
The x-axis is normalized coordinates  by the domain size $L$.
Results with our modified LB model are plotted as 'Present'  accompanied with analytic solutions and results with the LB models in Section~\ref{LB_model_pot_issue} with the central difference scheme .
In the Couette flow, velocity is assigned on the left wall (center) and right wall (right), respectively.\label{fig:Poise_Cou}}
  \end{center}
\centerline{\includegraphics[width=12cm]{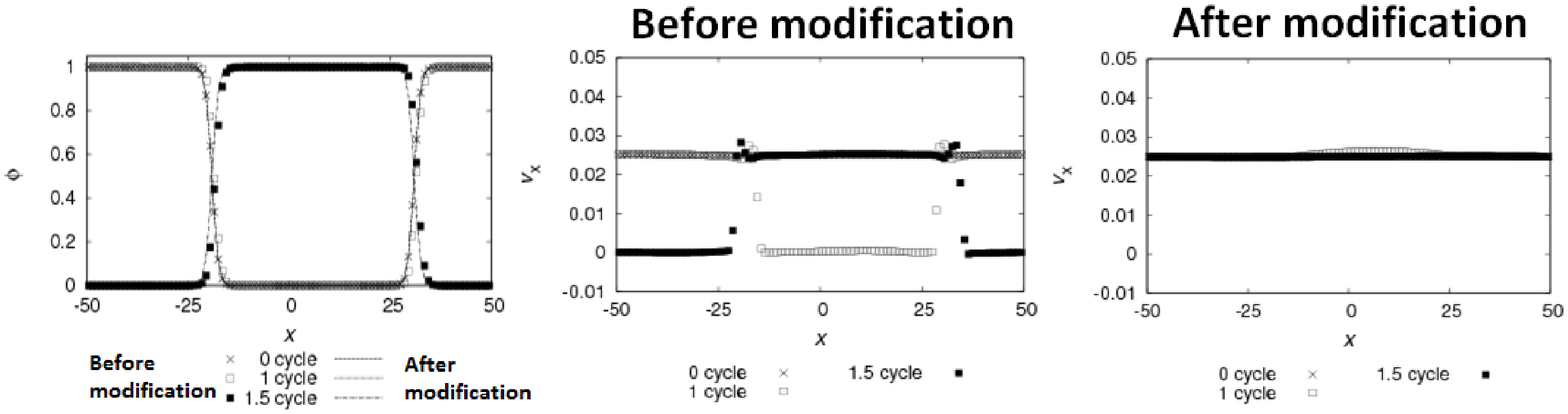}}
\caption{   Profiles of $\phi$ (left) and velocity $v_x$ with the original LB models (center) and the modified LB models (right) in terms of cyclic periods of droplet's movement. \label{fig:Galilean1D_dens}}
\vspace{0.1cm}
\centerline{\includegraphics[width=12cm]{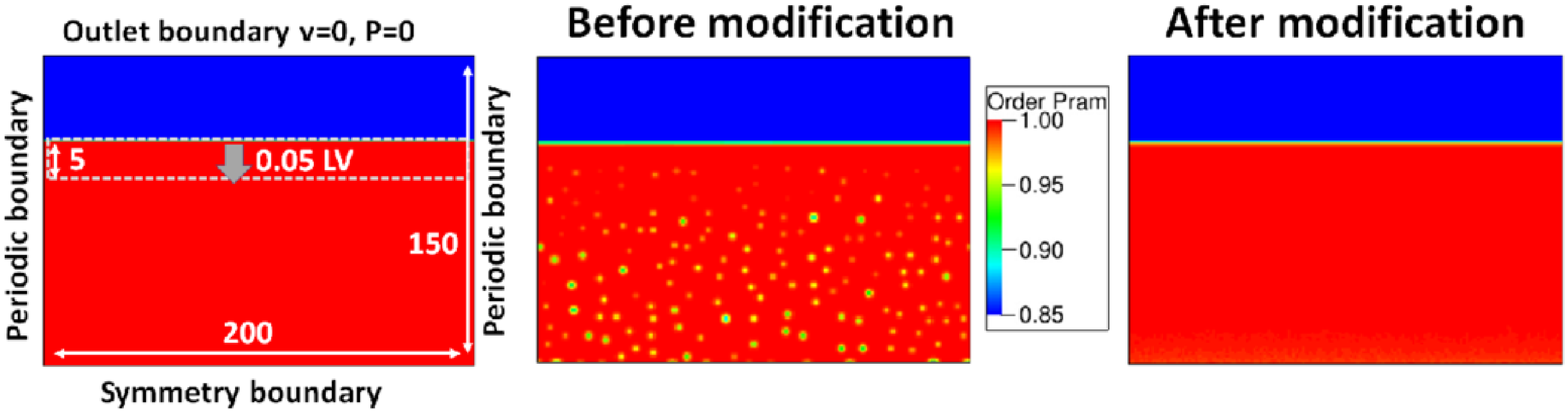}}
\caption{  Settings of an initial shock case (left). Its numerical results of $\phi$ ranging 0.85-1.0 at steady states with the original (center) and modified (right) LB models.\label{fig:Shock_dens_bubble} The central difference scheme is used for calculations of $\vec{n}$, $P$, $\rho \vec{u}$, and $K_i$. $M=1.7e-1$.}
\vspace{0.2cm}
\centerline{\includegraphics[width=13cm]{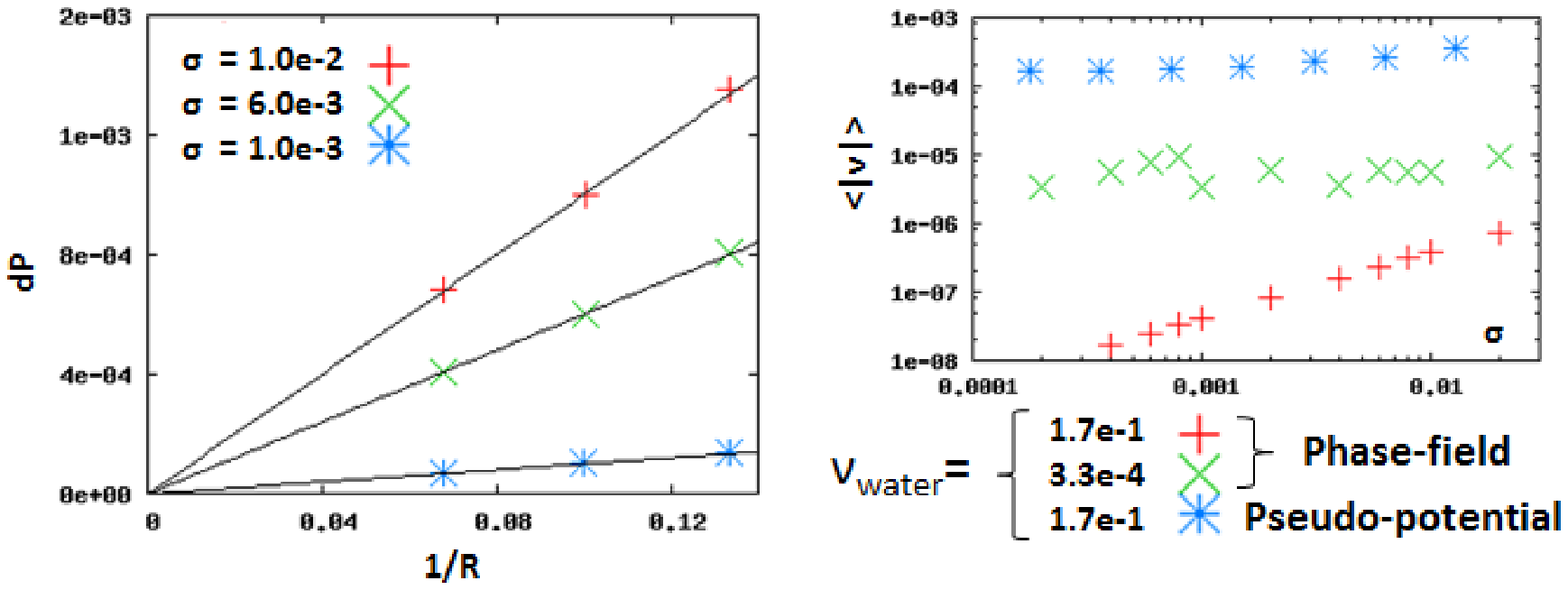}}
\caption{ Inverse droplet radius $1/R$ vs pressure difference $dP$ across the droplet interface (left).  
Surface tension $\sigma$ vs spatial averaged spurious current $<|v|>$ with $\nu_{water}= \left\{ 3.3e-4, 1.7e-1 \right\}$ (right).
Results are compared to a recent study of the pseudo-potential model for $\rho_{ratio} =1000$.\label{fig:2D_droplet_cntLap}}
\end{figure}

First, with the LB models in Section~\ref{LB_model_pot_issue}, shear stress on the interface is not evaluated accurately.
In Fig.~\ref{fig:Poise_Cou}, numerical results for two-phase Poiseuille flow and Couette flow are shown. In both cases, water  occupied the left half domain and air is on the right half side.
The domain sizes $L$ are 64 and 100 for each case. 
$\mu_{ratio}=100$, $\nu_{water}=0.17$, $M=0.1$, and $W=2.5$ for both cases. Gravitational acceleration $g=1.0e-6$  is applied in the Poiseuille flow.
The analytic solutions are derived while density profiles with non-zero interface thickness and viscosity in the mixture are taken into  account.

Results of the LB models in Section~\ref{LB_model_pot_issue} show peaks on the interface and deviate from analytic solutions obviously.
Since the similar behavior is observed even with the MRT collision operator in the previous study~\cite{2017_Fakhari_JCP} if they use the pure CD scheme, it is not mainly due to the filtered collision operator but likely due to the CD scheme.
Indeed, our analysis reveals that irregular effects at the interface originated from some high-order terms such as $\partial_{x} \left( v_y \partial^3_{x} \rho \right)$ in the Navier-Stokes (NS) equation where the CD scheme is used.
Although using the biased scheme can solve the issue as the
study~\cite{2017_Fakhari} indicated, such an approach requires information from
sites farther than the nearest neighbors, resulting in computational complexities.
Instead, a high order correction in the equilibrium distribution is added as the following in the present work together with the CD scheme and modification of definition of velocity so that the term $\partial_{x} \left( v_y \partial^3_{x} \rho \right)$ is removed,
\begin{eqnarray}
\delta f^{eq}_i = \frac{1}{4T}  w_i \left( \vec{c}_i  \cdot \vec{u} \right) \vec{\nabla}^2 \rho.
\end{eqnarray}
Fig.~\ref{fig:Poise_Cou} shows results of the modified model marked as ``Present'' matching with analytic solutions very well.

It is worth mentioning that this issue may be sometimes hidden in pressure or
force driven flow \cite{2017_Liang} because the higher density side has relatively lower velocity.
Due to the same reason, if velocity is assigned on a wall of the air side in the Couette flow, the velocity with the original models apparently matches with the analytic solution as shown in the right figure of Fig.~\ref{fig:Poise_Cou}.

Here we would like to emphasize that the purpose of comparisons in Fig.~\ref{fig:Poise_Cou} is to check the local balance of shear stress in each lattice by taking account of the non-zero interface thickness and fixing the interpolation of viscosity on the interface. 
This is different from a recent paper \cite{2017_Fakhari}, in which the interpolation method of $\tau_{mix}$ is studied. 
In our study, if the formula in Eq.~(\ref{tau_interpolate}) is changed, both of analytic solutions and numerical results are changed in Fig.~\ref{fig:Poise_Cou}.

In the previous study~\cite{2017_Fakhari}, this issue was improved with the so-called
velocity-based LB model, in which the first moment of the distribution function
is velocity instead of momentum. Therefore, in contrast to the momentum-based LB
model used in the present work, density and velocity are conserved independently, but there
is no guarantee of momentum conservation during the particle advection even at
zero surface tension.

\

Furthermore, this modified LB model also improves the Galilean invariance.
In the one-dimensional domain of $L$=100 bounded by periodic boundaries, the droplet is sitting in the center and a value of 0.025 for lattice velocity is homogeneously assigned initially.
$\nu_{water}=1.7e-2$ and $W=3.0$. The rest of settings are the same as in the above cases of Couette flow.
In Fig.~\ref{fig:Galilean1D_dens}, profiles of $\phi$ and velocity $v_x$ are presented in terms of the cycle period during which the droplet comes back to the original position. 
As seen here, though droplets' movements are similar with both models, the velocity with the modified model maintains a constant value and preserves Galilean invariance during the entire evolution, while the velocity with the original model described in Section~\ref{LB_model_pot_issue} fails.

\

Second, with the LB models for $\phi$ from Section~\ref{LB_model_pot_issue}, a large number of unphysical droplets/bubbles whose $\phi$ is close to 0/1, can be produced. 
In order to clarify this issue, a simple two-dimensional case is set as shown in the left figure of Fig.~\ref{fig:Shock_dens_bubble}. 
Here, the lattice velocity of 0.05 is assigned in the thin layer of water domain. $\mu_{ratio}=1$, $\nu_{water}=1.7e-1$, $W=3.0$, and $M=1.7e-1$. 
The shock created by initial velocity is reflected on the top and bottom edges of domain and bounces back and forth for a while. 
It produces a lot of bubbles which stay in the water domain even after the steady state is reached as shown in the center figure of Fig.~\ref{fig:Shock_dens_bubble}. 
Because of the steady state, it is likely that the value of the  mobility does not depend on the solution of these bubbles in the AC equation.
As the contour's range indicates, these are dense bubbles, whose $\phi$ is $0.90-0.99$. 
They are totally different from the "decent" bubbles, whose $\phi$ goes down to zero in their centers.
Hence such dense bubbles are insensitive to buoyancy force and sometimes distort streamlines.
This issue seems to be originated from the Allen-Cahn model, that prompts the nucleation of droplets and bubbles no matter how much $\phi$ they have.
In order to solve this problem, a minor diffusion correction is added to the Allen-Cahn model via $\theta$ in Eq.~(\ref{theta_phai_LBeq}) so that such bubbles are diffused. 
Specifically, $\theta$ is corrected as,
\begin{eqnarray}
\delta \theta = C \frac{M}{T} F \left( \phi, \vec{\nabla}^2 \phi \right) \mid \vec{\nabla} \phi  \mid,
\end{eqnarray}
where $C$ is a constant value. There are various choices of function $F$ but it is determined so that this correction is turned off for the decent bubbles and droplets.   
As a result, dense bubbles disappear as seen in the right figure of Fig.~\ref{fig:Shock_dens_bubble}.
In the test cases discussed in the next section, no obvious issues resulting from this modification are observed.
It indicates that the main interface dynamics is insensitive to this minor diffusion correction.
The similar problem of rarefied droplets in the air domain can be improved with a similar modification.

%%%%%%%%%%%%%%%%%%%%%%%%%%%%%%%%%%%%%%%%%%%%%%%%%%%%%%%
%%%%%%%%%%%%%%%%%%%%%%%%%%%%%%%%%%%%%%%%%%%%%%%%%%%%%%%
%%%%%%%%%%%%%%%%%%%%%%%%%%%%%%%%%%%%%%%%%%%%%%%%%%%%%%%

\section{A set of regular validation cases}

\subsection{A static droplet}
% motivation
Through the simulation of a static droplet in free space, consistence with the Laplace law is examined and the spurious current is  compared to the other multi-phase LB models. 
%
% case set up
A two-dimensional static droplet with variable initial radius, $R= \left\{ 8,
12, 16 \right\}$, is put in the center of domain, whose size is five times
of $R$ and periodic boundaries are assigned on each pair of domain's edges.
$\nu_{water}= \left\{ 3.3e-4, 1.7e-1 \right\}$, $\mu_{ratio}=60$, $M=0.1$, and $W=2.5$.
%
% result & discussion
First, with a small viscosity of $\nu_{water}=3.3e-4$,  a droplet with $\sigma = \left\{ 1.0e-2, 6.0e-3, 1.0e-3 \right\}$ is simulated.
The left figure of Fig.~\ref{fig:2D_droplet_cntLap} shows the relation between $1/R$ and the pressure difference across the interface, $dP$. 
Lines with slopes of inputted $\sigma$ are presented. 
All cases comply with the Laplace law, $dP=\sigma /R$, very well and output the consistent value of $\sigma$.
Next, spatially-averaged spurious current of a droplet of $R$=40 is measured for various $\sigma$ in the periodic domain of $250 \times 250$. 
In the right figure of Fig.~\ref{fig:2D_droplet_cntLap}, results with $\nu_{water}= \left\{ 1.7e-1, 3.3e-4 \right\}$ are compared to the previous study \cite{2016_Lycett} in which the recent pseudo-potential model for $\rho_{ratio}=1000$ is used.
While more diffusion with higher viscosity leads to less spurious current, the difference between two viscosity options becomes small as $\sigma$ is increased. 
The phase-field LB model in this study shows much improved spurious current than the recent pseudo-potential model by factor of $10^3 - 10^4$ for this droplet case of $\rho_{ratio}=1000$.

\subsection{Droplet collision}
% motivation
Binary droplet collisions are often simulated in order to check the interface dynamics and robustness of computational models for multi-phase flows with high-density ratio \cite{2016_Lycett,2004_Inamuro,2016_Inamuro}. 
In this study, a binary droplet collision under two flow conditions are compared to experimental results \cite{2009_Pan}.
%
% case set up
For each case, the droplet diameter is $\left\{ 32, 50 \right\}$, $\sigma= \left\{ 5.5, 0.64 \right\}$, and $\nu_{water}= \left\{ 1.9e-3, 6.0e-4 \right\}$.  $\nu_{ratio}=60$, $M=0.1$ and $W=2.5$.
The relative velocity $U$ is set as $0.1$ and $0.75$ for the case with coarse and fine resolution, respectively.
Hence Reynold number $Re$, $UD/ \nu_{water}$, is $\left\{ 1700, 6300 \right\}$ and Weber number $We$, $\rho_{water} U^2 D / \sigma$, is $\left\{ 58, 440 \right\}$, respectively.
%
% result & discussion
As shown in Fig.~\ref{fig:Droplet_collisionII}, the simulated droplets' deformation and splashing patterns depending on $Re$ and $We$ seem to be comparable with Fig.~2 and Fig.~4(a) in the paper\cite{2009_Pan}.
Compared to a previous study with the recent pseudo-potential LB model for $\rho_{ratio}=1000$ \cite{2016_Lycett}, similar accuracy can be achieved using half interface thickness and quarter resolution approximately in this study.

\subsection{Rayleigh-Taylor instability}
\begin{figure}[htbp]
\centerline{\includegraphics[width=14cm]{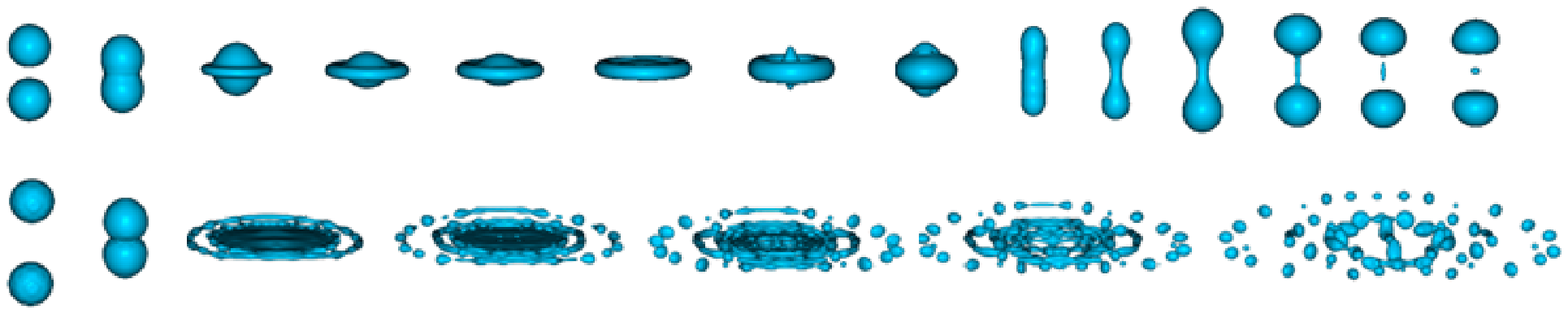}}
%\vspace*{8pt}
\caption{ Snapshots of a binary droplet collision. Reynolds numbers are 1700 (top) and 6300 (bottom). Weber numbers are 58 (top) and 440 (bottom).\label{fig:Droplet_collisionII}}
\vspace{0.5cm}
\centerline{\includegraphics[width=13cm]{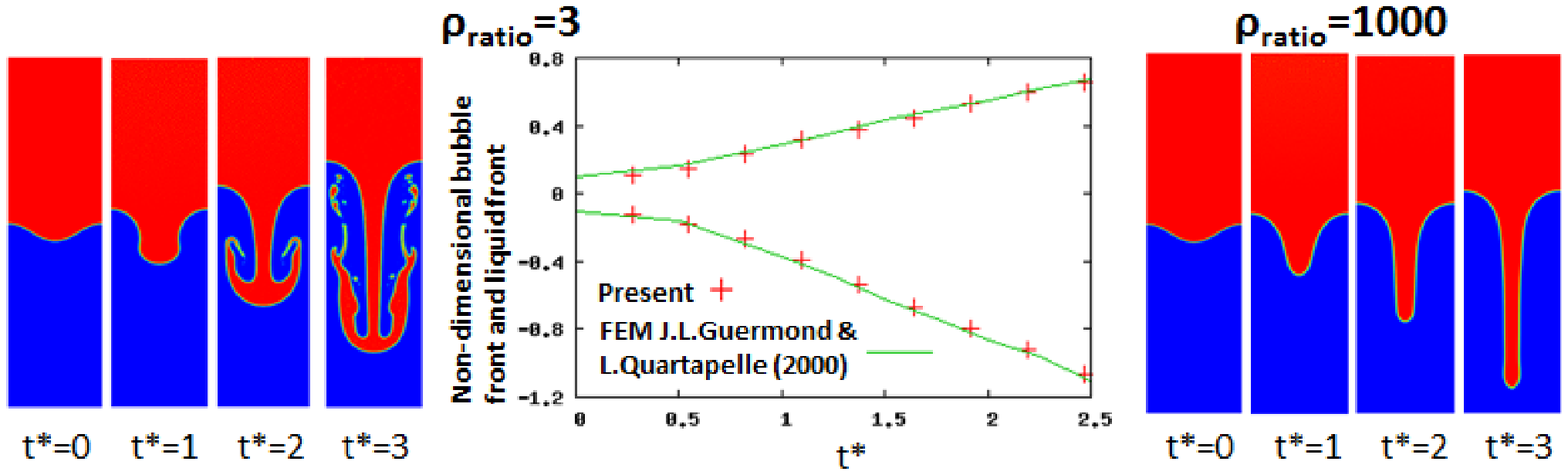}}
\caption{ Contours of $\phi$ where $\rho_{ratio}=$3 (left) and
 1000 (right), and histories of bubble/liquid front normalized by $L$ where $\rho_{ratio}=$3 (center) in Rayleigh Taylor instability. In the center figure, results of a previous study with the finite element method (FEM) are compared. $t^{*}=t  \sqrt{g/L}$. \label{fig:Ray_tay_inst}}
\vspace{0.5cm}
\centerline{\includegraphics[width=13cm]{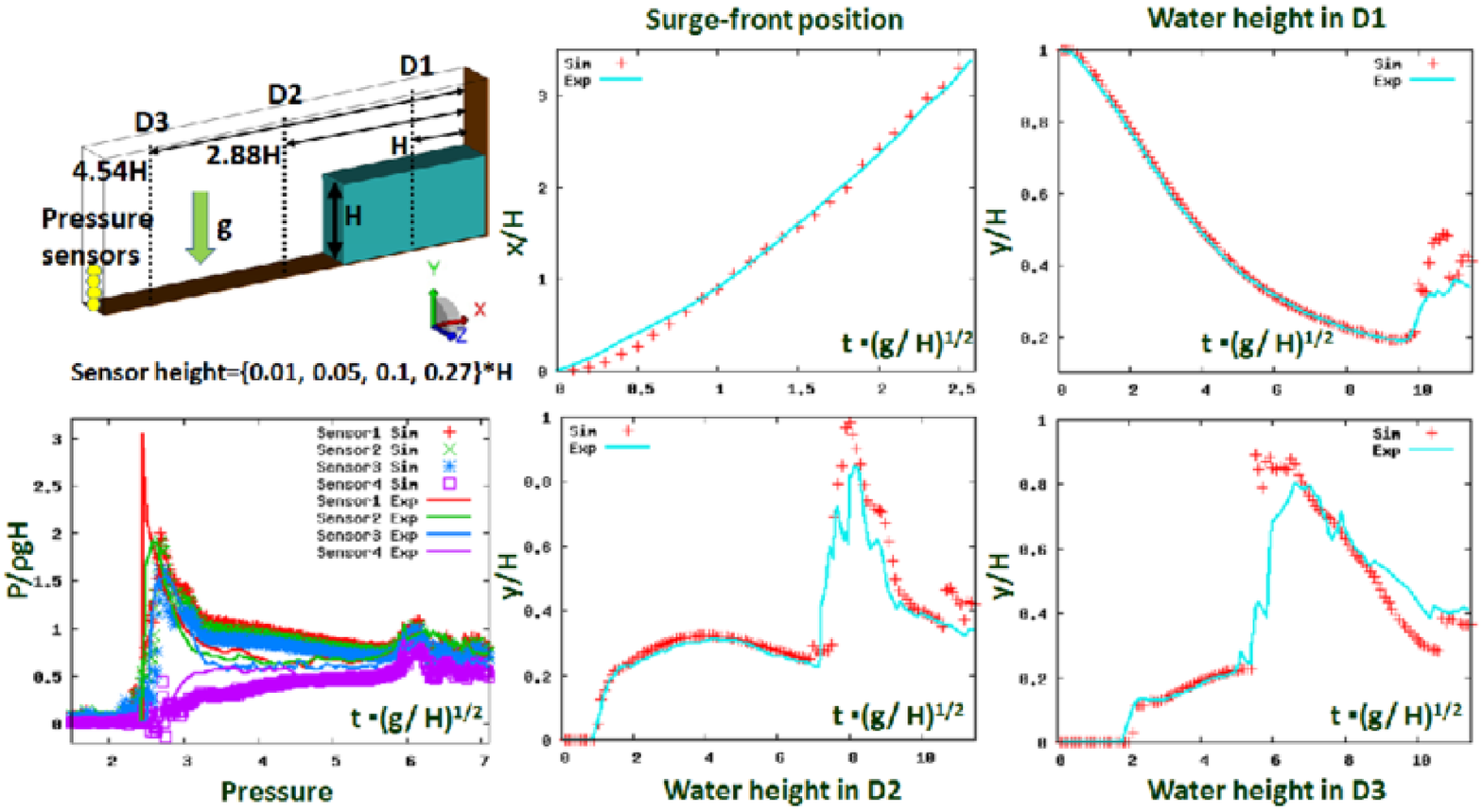}}
\caption{ In the dam-breaking case, settings (left top), comparisons of the surge-front position (center top), pressure on the wall  (left bottom), and water heights (others) between the simulation displayed as 'Sim' and the experiment displayed as 'Exp' \label{fig:Dam_breaking}}
\end{figure}

% motivation
Rayleigh-Taylor instability induced by heavy fluid's penetration into light fluid with gravity is simulated as a benchmark problem for the interface dynamics. Numerical results are compared to previous studies \cite{2017_Fakhari,2000_Guermond}.
%
% case set up
Two cases using $\rho_{ratio}= \left\{ 3, 1000 \right\}$ are tested in two-dimensional domains of $\left\{ 150 \times 600, 256 \times 1000 \right\}$, whose horizontal lengths are denoted as $L$. 
For each case, $\sigma= \left\{ 8.6e-4, 4.7e-1 \right\}$, $\mu_{ratio}= \left\{ 1, 100 \right\}$, $W= \left\{ 2.5, 5.0 \right\}$, $ M = \left\{ 5.8e-3, 1.3e-2 \right\}$, and $\nu_{water} = \left\{ 1.9e-3, 4.1e-3 \right\}$, respectively. 
Gravitational acceleration $g$ is set as $1.e-5$ for both cases. Accordingly, if the characteristic velocity $U$ is estimated as $\sqrt{g L}$, $Re $ = $UL/\nu_{water}$= 3000 and  the P\'{e}clet number, $UL/M$, is 1000 approximately. Capillary number, $\rho_{water} U \nu_{water} / \sigma$, is $\left\{0.26, 0.44 \right\}$, respectively.
%
% result & discussion
In the left and center figures of Fig.~\ref{fig:Ray_tay_inst}  contours of $\phi$ and histories of  bubble/liquid front normalized by $L$ where $\rho_{ratio}=3$ are presented.
The non-dimensional time $t^{*}$ is defined as $t  \sqrt{g/L}$.
The penetration pattern and its quantitative positions are consistent with previous studies\cite{2000_Guermond}, in which a finite element method was used.
In the right figure of Fig.~\ref{fig:Ray_tay_inst}, contours of $\phi$ with $\rho_{ratio}=1000$ are shown. 
Compared to results of the previous study\cite{2017_Fakhari} with the velocity-based phase-field LB models, the same-level robust results are captured while $W$ seems to be smaller and the finger shape is slightly sharper in the present study.

\subsection{Dam-breaking case}
% motivation
In a recent study \cite{2014_Lobovsky}, dynamics of the dam-breaking wave is investigated experimentally by measuring the water heights and the impact pressure on the downstream vertical wall.
This case is simulated with the LB models in this study and measurements are compared with the experiment.

% case set up
The computational domain of $537 \times 198 \times 50$ is bounded by friction walls except for the top boundary of the pressure condition. 
The initial height of water column $H$ is 100.
$g=4.0e-6$. $W=4.0$, $M=0.02$, and $\nu_{air}=8.3e-3$.
$\sigma = 5.2e-3$ and $\nu_{water}=4.9e-6$ so that the Weber number is $7.7e3$ and the Reynolds number of water is $4.1e5$.
The sensors' positions for the pressure measurement are shown in Fig.~\ref{fig:Dam_breaking}.
The water heights are detected on three lines, D1 through D3, located in the center of the z-coordinate.

% result & discussion
In Fig.~\ref{fig:Dam_breaking}, the measured pressure, the surge-front position of water, and water heights are compared with the experimental results.
Results for pressure, front position, and water height are nearly consistent with observations.
We note that although our proposed model lacks treatment for turbulence and the simulations are underresolved for the Reynolds number used, the physical structures captured by our model are nearly consistent with experiments.
In the beginning stage, the surge-front position is slightly underestimated due to the initial condition.
We believe the prediction of the pressure peak may be improved with increased spatial resolution.

\section{Summary}

An improved phase-field lattice Boltzmann model based on the Allen-Cahn equation with the filtered collision operator and the equilibrium distribution with a high-order correction in the LB equations for $\phi$ and hydrodynamic quantities is introduced without sacrificing  the simplicity of algorithm and computational cost significantly. 
These modifications improve solver robustness at low viscosity, accuracy of shear stress at interfaces, the Galilean invariance, and production of artificial dense bubbles and rarefied droplets.
In addition to extremely low spurious current, the same accuracy as the recent pseudo-potential model is achieved using half  of the interface thickness and a quarter of resolution.
Across various cases, consistent results with other numerical methods, analytic solutions and experiments are obtained with the efficient choices of parameters such as the resolution and the interface thickness.

\section*{Acknowledgments}
H.O. would like to thank Takaji Inamuro and Taehun Lee for effective and enlightening discussions.
We would also like to thank Luca D'Alessio for discussing setups of test cases. Also, we express gratitude to Casey Bartlett and Ilya Staroselsky for valuable comments on this study.

\clearpage

%+++++ Figure (Galilean invariance, Shock_dense_bubble_setting, Droplet setting) ++++++
%+++++ Figure end ++++

\clearpage

%+++++ Figure (Droplet setting, Rayleigh_Taylor_instability, Dam_breking ) ++++++
%+++++ Figure end +++++

\end{document}